# MLA-BIN: Model-level Attention and Batch-instance Style Normalization for Domain Generalization of Federated Learning on Medical Image Segmentation

**Fubao Zhu**[1], **Yanhui Tian**[1], **Chuang Han**[1], **Yanting Li**[1], **Jiaofen Nan**[1], **Ni Yao**[1*] and **Weihua Zhou**[2*]

[1]School of Computer and Communication Engineering, Zhengzhou University of Light Industry, China
[2]Department of Applied Computing, Michigan Technological University, USA
fbzhu@zzuli.edu.cn, yanhuitian6@gmail.com, hanchuang@zzuli.edu.cn, ytli1227@126.com,
nanjiaofen@zzuli.edu.cn, yaoni@zzuli.edu.cn, whzhou@mtu.edu

## Abstract

The privacy protection mechanism of federated learning (FL) offers an effective solution for cross-center medical collaboration and data sharing. In multi-site medical image segmentation, each medical site serves as a client of FL, and its data naturally forms a domain. FL supplies the possibility to improve the performance of seen domains model. However, there is a problem of domain generalization (DG) in the actual deployment, that is, the performance of the model trained by FL in unseen domains will decrease. Hence, MLA-BIN is proposed to solve the DG of FL in this study. Specifically, the model-level attention module (MLA) and batch-instance style normalization (BIN) block were designed. The MLA represents the unseen domain as a linear combination of seen domain models. The attention mechanism is introduced for the weighting coefficient to obtain the optimal coefficient according to the similarity of inter-domain data features. MLA enables the global model to generalize to unseen domain. In the BIN block, batch normalization (BN) and instance normalization (IN) are combined to perform the shallow layers of the segmentation network for style normalization, solving the influence of inter-domain image style differences on DG. The extensive experimental results of two medical image segmentation tasks demonstrate that the proposed MLA-BIN outperforms state-of-the-art methods.

## 1 Introduction

In deep learning for medical images, there is a desire to use multi-center data to build more accurate prediction and diagnosis models while ensuring the protection of user privacy in cross-center medical collaboration [Lehmann *et al.*, 2021; Zhou *et al.*, 2020]. Federated learning (FL) [Konečný *et al.*, 2016] provides a balanced solution between these two requirements. FL consists of a server and multiple clients at different centers, as shown in Figure 1. During the training process, the server first distributes the global model to each client, and then each client uses its own center's data to train a local model. Afterwards, the locally trained models are uploaded to the server. Finally, the server aggregates these models and obtains a new global model to complete one round of training. In FL, each client independently collects, annotates, and uses local data, and datasets from different centers are defined as different domains. Therefore, a global model with cross-domain features and higher accuracy can be obtained through multiple rounds of iterative training.

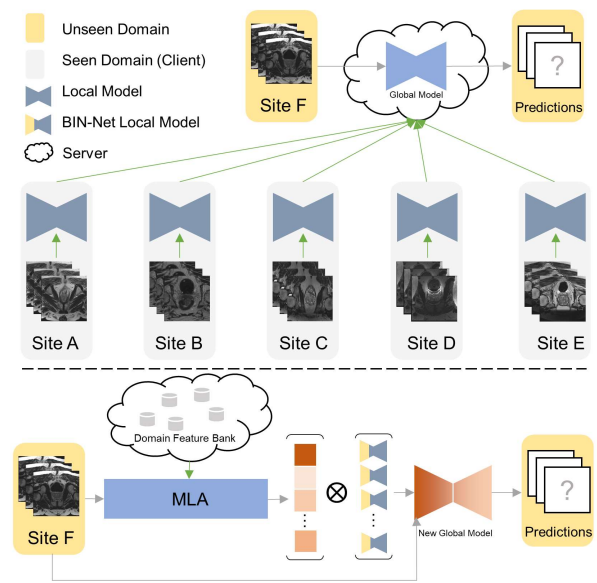

Figure 1: Illustration of (top) traditional FL training method and FL model testing on unseen center and (bottom) our proposed MLA-BIN model. The green arrow denotes the training phase, while the gray arrow denotes the testing phase.

Despite the effectiveness of FL in utilizing multi-center data [Zhang *et al.*, 2021; Yan *et al.*, 2021], current federated models underperform in the generalization of unseen domain data known as the domain generalization (DG) problem. This limitation is particularly evident in clinical deployment, where variations in imaging instruments and protocols in different medical centers lead to disparities in image data and features of convolutional neural networks (CNN) trained on data from different centers. These challenges make it difficult for the model to achieve good application performance in newly deployed institutions. The need for data annotation arises when the model cannot produce the desired segmentation in a new institution, a practice that is both time-consuming and impractical. The potential of FL in practical applications is limited by the low reusability of such models and their inability to be used for in-depth clinical research. Hence, solving the DG problem in FL is the key to realizing the practical applications of FL. It is also the fundamental purpose of promoting medical data collaboration and sharing.

Numerous studies have proposed effective solutions to enhance the generalization ability of models for resolving DG problems [Wang *et al.*, 2020; Gu *et al.*, 2021; Hu *et al.*, 2021; Pan *et al.*, 2018]. One of the most comprehensible techniques is to apply generative adversarial networks (GAN) for learning the stylistic features of images in the source domain and transferring style to the target domain [Yang *et al.*, 2018; Liu *et al.*, 2021]. Nevertheless, this approach could cause content information loss and interference in practical applications. Another conventional approach focuses on improving model robustness via extensive data augmentation [Zhang *et al.*, 2020]. However, it is necessary that the generalization ability in a fresh environment be upgraded through experience settings, thus rendering it less practical in real-life contexts. Novel adaptive solutions have attracted considerable research [Gu *et al.*, 2021; Hu *et al.*, 2021; Deng *et al.*, 2021; Wang *et al.*, 2020]. Nonetheless, the use of data from multiple centers is required during the training phase of these methods. Given that safeguarding data privacy is crucial for FL (i.e., data is distributed across different centers), such methods are infeasible and unsuitable for the FL paradigm.

Recently, novel DG solutions for FL have emerged which overcome the problem of data centralization [Liu *et al.*, 2021; Zhang *et al.*, 2021; Li *et al.*, 2021; Fallah *et al.*, 2020]. For feature differences, the current solutions involve utilizing GAN to align features from different domains [Zhang *et al.*, 2021] and applying contrastive learning to correct the local model feature deviations [Li *et al.*, 2021]. For style differences, the primary solutions comprise of enforcing style feature invariance [Liu *et al.*, 2021], data augmentation [Zhang *et al.*, 2020], and meta-learning paradigms [Fallah *et al.*, 2020]. However, these solutions solely focus on problem-solving, neglecting the practical deployment issues. Specifically, meta-learning and GAN are likely to impose a heavier client burden. This is especially true for real-world situations such as medical institutions, where the GPU configurations are limited, thus necessitating lightweight solutions.

To solve the DG problem in FL, we propose a novel lightweight approach, that adapts the deep and shallow layer features of images to the unseen domain. This approach aims to eliminate features and style differences present between seen and unseen domains. Specifically, a linear combination of a set of seen domain models is used to represent the global model of an unseen domain. During the testing phase in the unseen domain, the weights of each seen domain model are adaptively adjusted based on the similarity between the deep layers features of the unseen and seen domains. At the same time, the shallow layers features in the model are adjusted according to the feature of the network architecture to achieve style normalization of different domain images, as shown in Figure 1.

The motivation for our approach stems from the observation that FL learns cross-domain features by combining models from each client. However, using fixed weight parameters based on data quantity will lead to the global model being affected by the quantity of data in each client. For example, if the domain with a large number of images does not match the feature of images in the unseen domain, it will lead to the decrease in the performance of the global model in the unseen domain. Therefore, an attention mechanism that adaptively assigns weight parameters to each client based on the similarity of features between the unseen and seen domains is designed. In addition, following the findings of style transfer studies [Ulyanov *et al.*, 2016; Huang and Belongie, 2017], the characteristics of CNN architecture is to store style information in the statistical information of each feature channel in the shallow layers of the network (i.e., mean and variance). Instance Normalization (IN) can normalize the statistical information of each feature channel, while Batch Normalization (BN) can prevent internal covariate shift during the training process. Therefore, a block using both BN and IN in the shallow layers of the network is devised to improve the performance of the local model and eliminate the style differences between domains.

Our main contributions are as follows:
- We proposed a model-level attention module (MLA) that is suitable for FL. The MLA represents the unseen domain as a linear combination of seen domain models and adaptively adjusts the weights of seen domain models through an attention mechanism, thereby enabling generalization to the unseen domain.
- We designed a batch-instance style normalization block (BIN) that is suitable for FL to solve the domain gap caused by style differences. It is combined with the segmentation backbone network (BIN-Net), which can not only effectively learn intra-domain features but also eliminate inter-domain style differences without requiring access to data from other centers.
- We conducted extensive experiments and presented experimental results in two medical image segmentation tasks. Our proposed method outperformed state-of-the-art DG methods in FL. The effectiveness of our method was demonstrated through in-depth analytical experiments.

## 2 Methods

### 2.1 Problem Definition and Method Overview

Let $D = \{D_1, D_2, ..., D_k\}$ be a set of $k$ source domains, with an unseen domain denoted as $D_u$. Each domain represents a data center and serves as a client in the FL setting. The $t$-th image and label in the $D_n$ domain are denoted as $x_n^t$ and $y_n^t$, respectively. To enable FL to generalize to unseen domain, the MLA is proposed that utilizes image features to detect affiliated domain, as shown in Figure 2(a). By considering the similarity of image features between seen and unseen domains and learning cross-domain features, the MLA avoids negative transfer effect and achieves a global model with high segmentation performance in the unseen domain, as shown in Figure 2(b). Inspired by style transfer, the BIN block for the shallow layers of segmentation network is designed to enable style normalization of different center data, as shown in Figure 2(c).

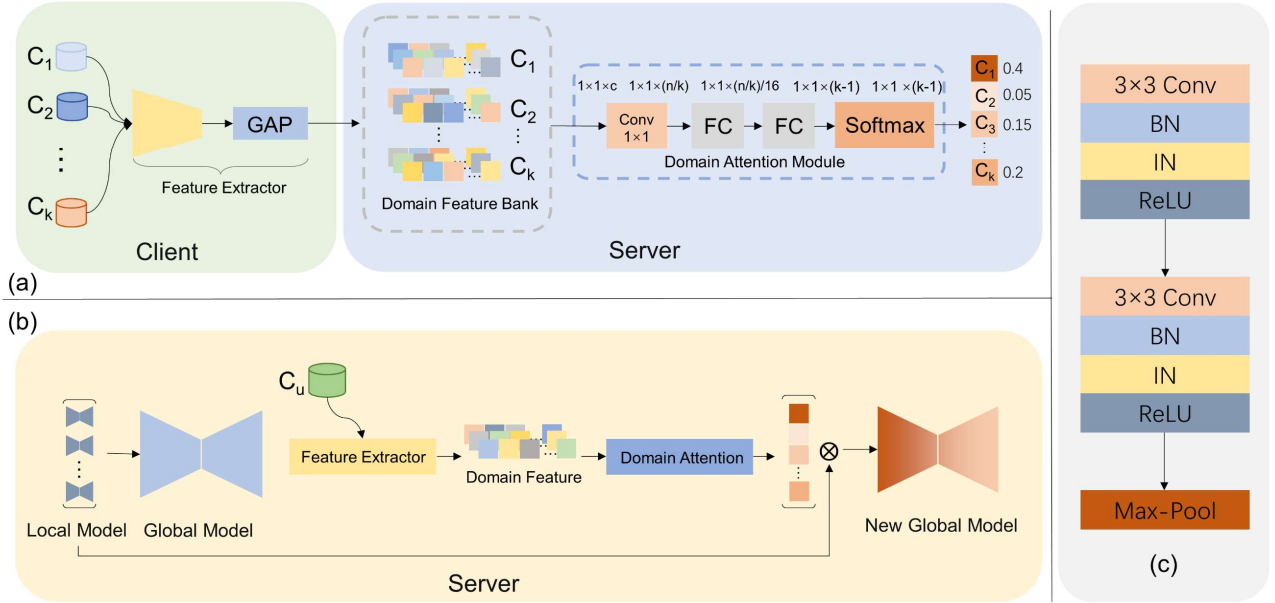

Figure 2: Overview of the proposed model-level attention (MLA) and batch-instance style normalization (BIN) block. (a) Details of the proposed MLA training phase. (b) Details of the proposed MLA application phase. (c) Structure of the BIN block.

## 2.2 Model-Level Attention

According to the definition of the aggregation phase in FL, the aggregation formula for the global model can be formally expressed as

$$G(m) = \sum_{i=0}^{k} w_i l_i(m), \quad (1)$$

where $l_i(m)$ represents the local model of the $D_i$ domain, $w_i$ represents the weight of the $D_i$ domain, and $G(m)$ represents the global model obtained by weighted summation of $k$ clients.

Therefore, aggregation can be expressed as a linear combination of models from seen domains to develop a global model using test data from unseen domain. The traditional aggregation methods use fixed values for weighting coefficients $w = \{w_1, w_2, ..., w_k\}$, such as the ratio of the center data quantity to the total data quantity. Nevertheless, in the case of test data from unseen domains, the difference in data features among distinct domains affects the weighting coefficient and results in sub-optimal global models. Therefore, an attention mechanism is introduced to obtain optimal weights of models based on the similarity of inter-domain data features and enable the global model to generalize to unseen domain.

In addition, traditional attention methods are designed for network-level attention by assigning weights to each feature channel in the network architecture. In contrast, MLA is more suitable for FL since it weights models of each domain instead of feature channels within the network architecture.

The MLA consists of three parts: feature extractor, domain feature bank, and domain attention module (DAM). The feature extractor is utilized to extract deep layer features from image data using CNN on the clients. The domain feature bank is used to store the image features on the server-side. The Domain Attention Module (DAM) is employed to identify the affiliated domain of the image features. The operational process of MLA will be elaborated below.

**Training**
MLA implements output at the model level, thus requiring a separately trained attention module as shown in Figure 2(a). The purpose of MLA is to identify similarities between input data and data from seen domains. Sharing seen domain images is prohibited in FL. Hence, it is proposed to use CNN as a feature extractor to extract deep features from images. Deep features cannot be restored to the original image, so sharing deep features does not result in leakage of real data. Specifically, a pre-trained CNN model with unified pre-training weights is first used to extract features from data in each client $D = \{D_1, D_2, ..., D_k\}$. Next, a global average pooling operation is performed on the extracted features, resulting in a mapping of images to the feature space. Finally, the extracted features from each client are uploaded to the server to construct a shared domain feature bank. The formula for mapping images to the feature space can be expressed as

$$C_n^t = GAP\big(Enc(x_n^t; \theta_{Enc})\big), \quad (2)$$

where $GAP$ represents global average pooling layer, $Enc$ represents the encoder part of the CNN with pre-trained model, $\theta_{Enc}$ represents the encoder parameters, and $C_n^t$ represents the representation of the $t$-th image in the $D_n$ domain mapped to the feature space.

The domain feature bank data is employed as the dataset to train MLA. These data are used as input features and the corresponding labels to train a domain-multilayer perceptron as DAM on the server. The DAM consists of a 1x1 convolutional layer and two fully connected layers. It outputs $k$ values through the soft-max function, denoted as $D^w$, where $k$ values correspond to the probabilities that the input features belong to $k$ domains. $D^w$ satisfies $\sum_{i=0}^{k} D_i^w = 1$. The formula for the DAM can be expressed as

$$D^w = SM\left(FC_2\left(FC_1\left(Conv(C_n^t; \theta_C); \theta_{FC_1}\right); \theta_{FC_2}\right)\right), \quad (3)$$

where $SM$, $FC$ and $Conv$ respectively represent the soft-max layer, fully connected layer, and convolutional layer. $\theta_C$ and $\theta_{FC}$ respectively represent the trainable parameters of the convolutional and fully connected layers.

**Application.** To test the model in an unseen domain on the server, the trained DAM is used to adjust the weight parameters of each central model to adapt the global model to the unseen domain, as shown in Figure 2(b). First, the data from the unseen domain $D_u$ is subjected to feature extraction using the same feature extractor used during MLA training and mapped to a feature space. Next, these features are used as inputs to the DAM for domain similarity prediction. The output of the module is the similarity between data features from the unseen domain and those from each seen domain. Subsequently, this output is used as aggregation weights to weight each client's local model, to obtain a new global model. Finally, this new global model is used to implement the application of MLA for inference on unseen domain data. The formula for the new global model can be expressed as

$$G(m) = \sum_{i=0}^{k} D_i^w l_i(m), \quad (4)$$

where $D_i^w$ represents the predicted probability of belonging to the $D_i$ domain.

### 2.3 Batch-Instance Style Normalization

**Background.** BN [He *et al.*, 2016] plays a crucial role in CNN by working on the network in batches, preventing gradient vanishing, accelerating training, and addressing the internal covariate shift problem. IN [Ulyanov *et al.*, 2016] is a deep learning technique used to normalize each instance in feature channels, which can realize normalization of fixed information of channels in the network. Fine-grained information, such as styles and colors that are fixed attributes in one image, are mainly processed by the shallow layers of CNN. Therefore, using IN in the shallow layers of the network can achieve image style normalization. The advantage of IN is that it can unify styles among several domains, even without other domain data distributions. Therefore, IN can be well applied in FL.

**Batch-Instance Normalization Block.** BN alone cannot achieve style normalization in image segmentation, while IN alone cannot fully exploit within-domain features. Both techniques are combined to take full advantage of their individual strengths, reducing intra-domain feature shift and achieving inter-domain style normalization. As shown in Figure 2(c), the premise of achieving style normalization is that the network performs well in their respective domains. Therefore, in the BIN block, a 3x3 convolution is first performed, followed by BN to ensure that the performance of the network within the domain is not diminished. Next, the feature channels style normalization is implemented by IN. Then, ReLU activation function is performed. And the whole process is repeated twice in each BIN block, followed by max-pooling operation. The formula for the BIN block can be expressed as

$$BIN_{sub} = Act\left(IN\left(BN(Conv(f; \theta_c); \theta_{BN})\right)\right), \quad (5)$$

$$BIN = MP\left(BIN_{sub_2}\left(BIN_{sub_1}(f; \theta_{sub_1}); \theta_{sub_2}\right)\right), \quad (6)$$

where $MP$, $Act$, $IN$, $BN$ and $Conv$ respectively represent max-pooling layer, ReLU activation function, IN layer, BN layer, convolutional layer. $f$, $\theta_C$, $\theta_{BN}$, $\theta_{sub}$ respectively represent block input, parameters of convolutional layer, parameters of BN layer and parameters of BIN$_{sub}$ block.

**Networks.** Previous research [Pan *et al.*, 2018] has suggested that style information for images is primarily located in the top-3 layers of networks. In the experimental section, it is demonstrated by extensive contrast experiments. Therefore, in this paper, the top-3 layers of the network are defined as the shallow layers. The efficient U-Net is selected as the backbone of segmentation network, with BIN block replacing convolutional blocks in the top-3 layers of the U-Net encoder, while retaining the original convolutional blocks in the fourth and fifth layers to construct the BIN-Net.

## 3 Experiments

The proposed MLA-BIN, a combination of MLA and BIN-Net, was extensively evaluated against baseline and state-of-the-art DG models of FL on two multicenter tasks. The tasks include segmentation of the prostate using MRI, segmentation of optic disc and cup (OC/OD) using color fundus images. The multicenter segmentation tasks cover different modalities of medical images and represent the DG challenges that exist in FL.

### 3.1 Dataset

Two datasets were employed in this study. For prostate segmentation, a public dataset consisting of 116 T2-weighted MRI cases from six different centers [Liu *et al.*, 2018], with case ratios of 30:30:19:13:12:12 for each center was used. To ensure consistency in evaluation, only MRI slices with prostate regions were retained for network training and testing. For OC/OD segmentation, a public dataset of 1070 retinal fundus images from four different centers was used [Wang *et al.*, 2018; Sivaswamy *et al.*, 2015; Fumero *et al.*, 2011; Orlando *et al.*, 2020], with case ratios of 101:159:400:400 for each center. Each retina image was cropped to a patch size of 384×384 as input. These datasets include images and pixel-level target area labels. All data were z-score normalized.

### 3.2 Implementation Details

During the FL training process, the same hyperparameters were used for model training across all clients. The batch size for the local models was set to 8, and Adam optimizer was used under default settings. The learning rate was set to 0.01, with a decay rate of 100 and a decay exponent of 0.8. Each local model was trained for 1000 epochs, and the global model was updated after each epoch to ensure the centralization of model parameters. For the segmentation networks of each client, the Dice loss is used for training, expressed as

$$L_{seg} = 1 - \frac{2|P \cap G|}{|P| + |G|}, \quad (7)$$

where $P$ represents the output of the model and $G$ represents the ground truth under one-hot encoding.

| Unseen Site | A | | B | | C | | D | | E | | F | | Average | |
|---|---|---|---|---|---|---|---|---|---|---|---|---|---|---|
| | DSC ↑ | ASD ↓ | DSC ↑ | ASD ↓ | DSC ↑ | ASD ↓ | DSC ↑ | ASD ↓ | DSC ↑ | ASD ↓ | DSC ↑ | ASD ↓ | DSC ↑ | ASD ↓ |
| BigAug | 89.73 | 1.77 | 85.84 | 2.14 | 84.29 | 2.27 | 88.29 | 1.33 | 83.15 | 3.23 | 89.27 | 1.12 | 86.76 | 1.98 |
| ELCFS | 90.62 | 1.16 | 86.49 | 1.53 | **85.43** | **1.59** | 89.73 | 1.08 | 83.28 | 3.18 | 89.18 | 1.04 | 87.46 | 1.60 |
| FedADG | 89.92 | 1.96 | 86.58 | 1.78 | 84.85 | 2.01 | 89.38 | 1.15 | 83.96 | **2.17** | 89.64 | 1.02 | 87.39 | 1.68 |
| DCAC | 90.56 | 0.95 | **89.54** | **1.16** | 84.76 | 2.16 | 89.41 | 1.12 | 83.54 | 2.49 | 89.36 | 1.25 | 87.91 | 1.53 |
| FedAvg | 88.36 | 2.12 | 84.80 | 2.31 | 83.48 | 2.58 | 86.42 | 2.61 | 79.85 | 4.49 | 88.22 | 1.87 | 85.19 | 2.66 |
| Ours (MLA-BIN) | **91.98** | **0.85** | 87.31 | 1.27 | 85.03 | 1.71 | **90.42** | **1.06** | **84.19** | 2.29 | **90.67** | **0.93** | **88.27** | **1.35** |

Table 1: Performance of our method, baseline model and four DG methods in prostate segmentation

| Unseen Site | A | | B | | C | | D | | Average | |
|---|---|---|---|---|---|---|---|---|---|---|
| | DSC ↑ | ASD ↓ | DSC ↑ | ASD ↓ | DSC ↑ | ASD ↓ | DSC ↑ | ASD ↓ | DSC ↑ | ASD ↓ |
| BigAug | (81.76, 93.68) | (19.61, 9.16) | (72.49, 86.73) | (25.34, 18.71) | (84.15, 93.42) | (11.24, 9.08) | (84.23, 95.42) | (8.74, 5.66) | 86.49 | 13.44 |
| ELCFS | (**84.75**, 94.26) | (**17.32**, 8.77) | (72.35, 86.40) | (25.57, 18.56) | (84.81, 94.26) | (10.73, 8.21) | (85.07, 96.42) | (8.12, 4.97) | 87.29 | 12.78 |
| FedADG | (82.96, 94.47) | (18.85, 8.65) | (73.49, 86.95) | (24.51, 17.93) | (84.29, 95.42) | (10.86, 7.87) | (85.28, 95.76) | (8.03, 5.59) | 87.33 | 12.79 |
| DCAC | (83.17, 94.72) | (18.73, 8.42) | (74.08, 87.66) | (23.27, 17.52) | (84.75, 95.75) | (10.44, 7.62) | (**86.89**, **96.71**) | (**7.51**, **4.88**) | 87.97 | 12.30 |
| FedAvg | (81.06, 92.90) | (20.42, 9.85) | (71.64, 84.26) | (26.19, 22.45) | (83.06, 93.12) | (11.75, 9.35) | (83.15, 95.57) | (9.23, 5.70) | 85.60 | 14.37 |
| Ours (MLA-BIN) | (83.49, **95.15**) | (17.81, **8.01**) | (**74.16**, **88.80**) | (**23.12**, **16.38**) | (**85.57**, **96.33**) | (**10.12**, **7.43**) | (86.29, 96.23) | (7.87, 5.13) | **88.25** | **11.98** |

Table 2: Performance (OC, OD) of our method, baseline model and four DG methods in OC/OD segmentation

The DAM of MLA used the same settings as FL, except that the learning rate was set to 0.001 and the epoch was set to 100. The DAM in MLA is a neural network consisting of multiple perceptron that is used for classifying image features to identify the domain to which the image belongs. This classification task is trained using the cross-entropy loss function, expressed as

$$L_{cls} = -\sum_{i=0}^{k} d_i \log(d_i^p), \quad (8)$$

where $d_i$ represents the label of the domain to which an image belongs, and $d_i^p$ represents the probability predicted by the network that the image belongs to that domain.

Our method was implemented using the PyTorch and trained on a workstation with two NVIDIA V100 GPUs.

### 3.3 Comparative Experiments and Analysis

The proposed method was compared with four state-of-the-art DG methods, including (1) BigAug method [Zhang et al., 2020] based on extensive data augmentation, (2) ELCFS method [Liu et al., 2021] based on the principle of amplitude spectrum and meta-learning, (3) FedADG method [Zhang et al., 2021] based on federated adversarial learning, and (4) DCAC method [Hu et al., 2021] based on domain and content adaptive convolution. Except for the DCAC method, all other methods used the U-Net as the backbone for model training in the FL framework. The baseline model was set up using the FedAvg [McMahan et al., 2017] algorithm, and it was compared with other methods.

For each segmentation task, datasets from k+1 centers were used, among which each dataset was used in turn as the unseen domain for performance evaluation. The remaining k datasets were used as seen domains for model training. By looping k rounds, each dataset could be evaluated once as the unseen domain. The segmentation results were evaluated using the Dice Similarity Coefficient (DSC) and Average Surface Distance (ASD). DSC (%) and HD (pixels) were used to assess the accuracy of the content and boundary segmentation results, respectively.

**Comparative Results in Prostate Segmentation.** Table 1 presents the quantitative results of DG for prostate segmentation task under the FL paradigm. The results indicate that although the BigAug method performs better than the baseline model, its performance is still inferior to other DG methods. It suggests that data augmentation methods are undirected and cannot accurately generalize data to unseen domains. The ELCFS method outperforms the FedADG method, but falls slightly short compared to the DCAC method, indicating that while the method of frequency space interpolation extends the distribution of the local dataset, it still lacks adaptability in unseen domains. Our proposed method achieves an average Dice improvement of 3.08% over the baseline model. Although our method performs slightly lower than other methods on individual domains, compared to the DCAC method that requires data sharing, our method has advantages in privacy protection. More importantly, compared to four state-of-the-art DG methods, our proposed method has better average Dice and average ASD performance. It indicates that our method can adapt unseen domain images to seen domain models, while eliminating style differences between different domains and improving the model's generalization ability in unseen domains.

**Comparative Results in OC/OD Segmentation.** Table 2 presents the quantitative results of DG for OC/OD segmentation under the FL paradigm. For the OC/OD segmentation task, the use of undirected data augmentation methods BigAug did not result in significant improvements. In contrast, the FedADG method performed slightly better than the BigAug and ELCFS methods, reflecting the advantages of using GAN for feature distribution alignment. The ELCFS method performed well on certain central data, possibly because the generated data is similar to the data in the unseen domain, allowing the model to better adapt in the unseen domain. Compared to these methods, our method has better adaptability, even in cases of limited data quality, and can generalize well to unseen domain.

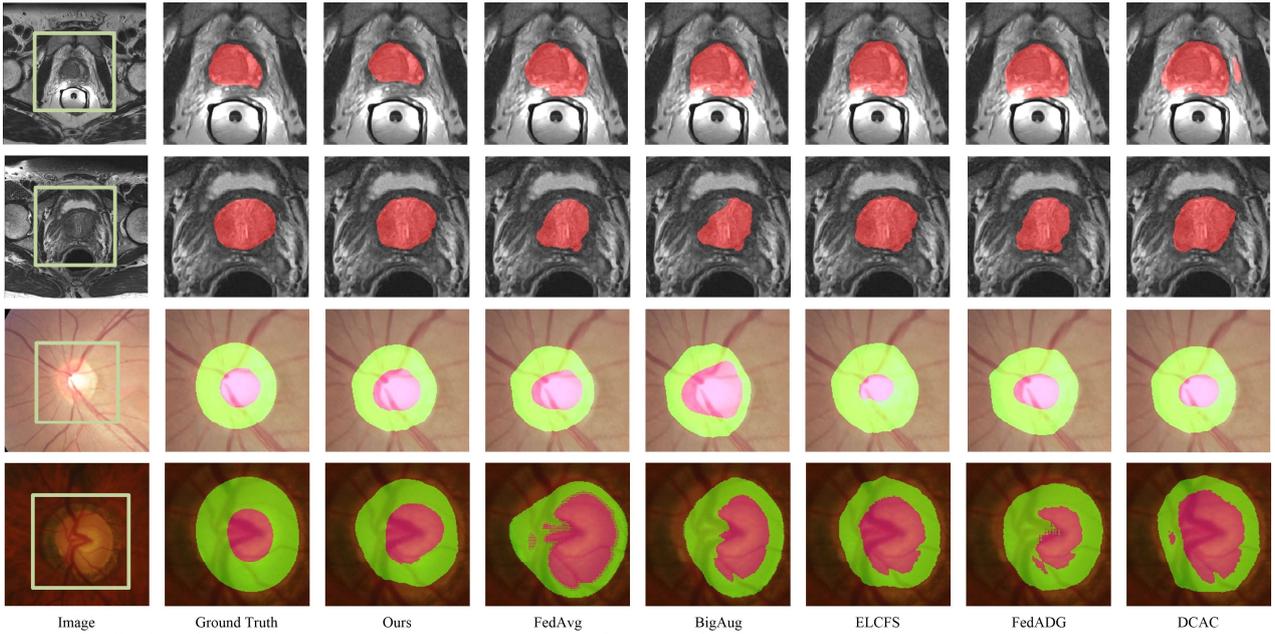

Figure 3: Visualization of the results predicted by ours (MLA-BIN) and four state-of-the-art methods on the two segmentation tasks, together with ground truth and baseline.

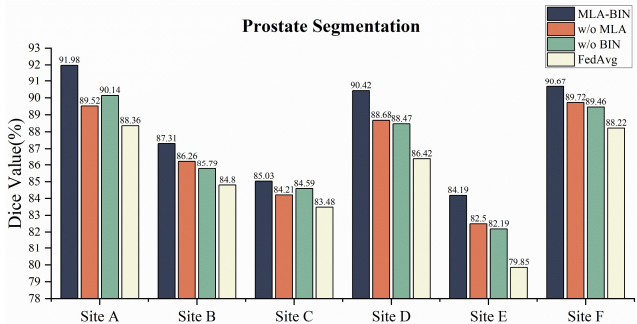

Figure 4: Ablation results to analyze the effect of the two components (i.e., MLA and BIN) in our method, together with baseline (FedAvg).

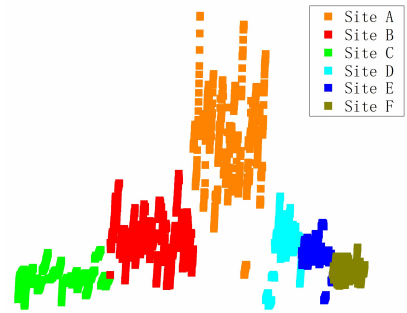

Figure 5: Visualization of the results of MRI prostate data from six sites using t-SNE dimensionality reduction.

**Visualization Results.** In Figure 3, a qualitative analysis was conducted to compare the DG performance of our method and four state-of-the-art methods on different tasks. Regardless of content segmentation or boundary segmentation, the results of each segmentation task showed that our method produces more accurate segmentation results. It confirms the effectiveness of our method in DG of FL.

**Ablation Analysis.** In this study, MLA and BIN modules were designed to enable the model to adapt to unseen domain data. To evaluate the contributions of these two modules, in MRI prostate segmentation the model performance is compared with only MLA, only BIN, and both modules used simultaneously, using FedAvg as the baseline model. As shown in Figure 4, both MLA and BIN modules can improve the generalization performance on the unseen domain. It indicates that there are domain feature shifts and style differences in DG, and MLA and BIN modules respectively address these two issues to improve the model's generalization performance. Simultaneously using MLA and BIN modules can compensate for the shortcomings of using a single component and further improve the model's generalization performance.

**Validation of data validity.** In the DG problem, the data in different domains exhibit non-independently identically distribution (non-IID) characteristic. The data utilized in this study are categorized into different domains based on various institutions, thus, the data naturally conforms to non-IID characteristic. Additionally, the t-SNE dimensionality reduction technique is employed to validate the presence of non-IID characteristics in this study. Figure 5 shows the visualization of results of six domains of MRI prostate segmentation data in dimensionality reduction to 2D space. The data from different domains exhibit distinct distributions in the 2D space, indicating the non-IID characteristic of the data. The data is suitable for DG problem.

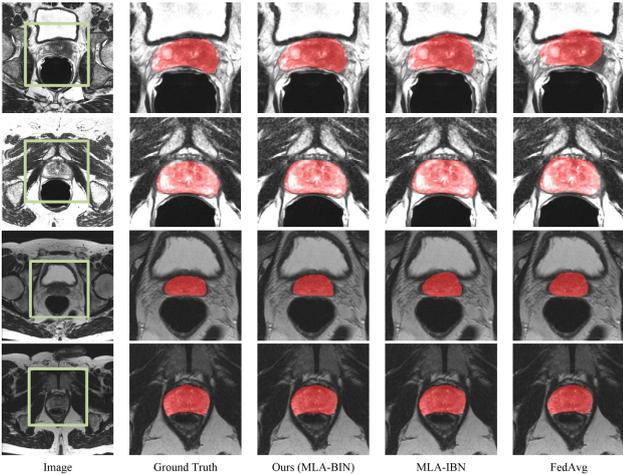

Figure 6: Visualization of four prostate MRI scans, the corresponding segmentation ground truth, and the results of our MLA-BIN, MLA-IBN, and baseline.

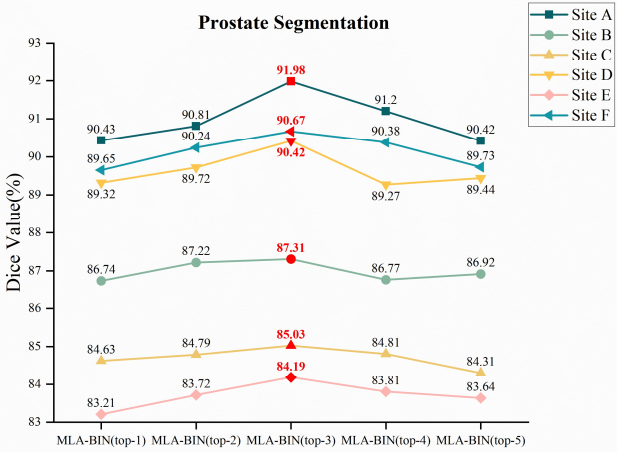

Figure 7: Performance of MLA with BIN in top-1 layer to top-5 layers on prostate segmentation task.

**BIN or IBN.** The BIN module proposed in this paper has a similar structure to the IBN module proposed by [Pan et al., 2018]. To better evaluate these two modules, the BIN module and IBN module were respectively used to train the model and conduct a qualitative analysis in MRI prostate segmentation. As shown in Figure 6, both MLA-BIN and MLA-IBN achieved reliable results in the unseen domain. However, MLA-BIN was superior to MLA-IBN in the details of segmentation, which indicates that the IBN module's strategy of concatenating IN and BN layers to solve style normalization can lead to incomplete style normalization, with the style differences that BN layers retain being transmitted to the next convolutional layer as features. The BIN module avoids this situation by adopting a strategy of linking BN and IN layers. Using IN after BN not only ensures the effective learning of intra-domain features but also fully achieves style normalization, thus avoiding inter-domain style differences.

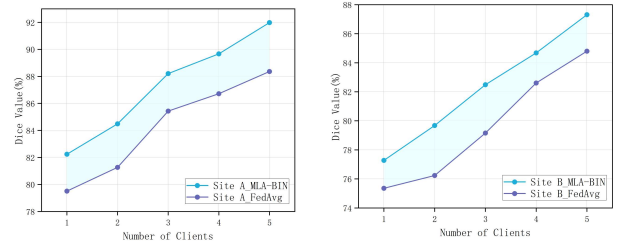

Figure 8: Generalization performance of models trained with different numbers of client sites with Site A and Site B serving as unseen domains, using MLA-BIN and FedAvg.

**Definition of shallow layer of network.** In [Pan et al., 2018]'s study, the shallow layers of the network were defined as the top-3 layers using feature divergence analysis. In this work, the point is corroborated from a different perspective by investigating the generalization performance of models with a BIN module embedded into the network from the top-1 layer to the top-5 layers, to confirm the definition of the shallow layers of the network. As shown in Figure 7, in MRI prostate segmentation, the best generalization performance is achieved when the BIN module is applied to the top-3 layers, while using the BIN module excessively or insufficiently results in weaker generalization performance than that of embedding the module in the first three layers. A possible reason is that using the BIN module only in the top-1 layer or top2 layers leads to incomplete normalization of image style differences across domains. Moreover, adding the BIN module on the basis of top-3 layers results in the filtration of deep features in the network, which is not conducive to complementary learning with MLA. Therefore, defining the shallow layers of the network as the top-3 layers and applying the BIN module for style normalization in these layers is optimal.

**Impact of the number of clients.** In addition, 1 to K-1 domains are randomly selected as training clients in seen domains, and the impact of different numbers of clients on the generalization of FL is analyzed. Figure 8 shows the generalization performance of Site A and Site B for the prostate segmentation task, using the MLA-BIN and FedAvg methods, with varying numbers of clients. With the increasing number of clients, the performance of MLA-BIN and FedAvg gradually improves. The improvement can be attributed to the increased number of clients, which enables a more comprehensive and accurate representation of known domains in relation to unseen domain. Furthermore, regardless of the number of clients used, MLA-BIN consistently outperforms FedAvg in terms of generalization performance, which further shows the effectiveness of MLA-BIN method in the DG of FL.

## 4 Conclusions

This paper proposes a novel method to address DG problems in FL by incorporating a model-level attention module (MLA) with batch-instance style normalization module (BIN). The MLA represents an unseen domain as a linear combination of seen domain models and adaptively adjusts the weights of seen domain models based on the similarity of deep layer features between unseen and seen domains. The BIN, which combines BN and IN, is applied to the shallow layers of the network to achieve effective learning of intra-domain features and inter-domain style unification. Our results in two multi-center medical image segmentation tasks demonstrate that our proposed method can effectively generalize from seen to unseen domains in FL paradigm and outperforms state-of-the-art DG methods.

## Acknowledgments

This research was supported by the Henan Science and Technology Development Plan (Project Number: 222102210219, 232102210010, 232102210062), Young Teacher Foundation of Henan Province (Grant No.2021GGJS093), the National Outstanding Youth Science Fund Project of National Natural Science Foundation of China under Grant 62106233, and the Doctor Natural Science Foundation of Zhengzhou University of Light Industry (2022BSJJZK13).

## References

[Lehmann *et al.*, 2021] Nicolas J. Lehmann, Muhammed-Ugur Karagülle, Felix Spielmann, Bianca George, Benjamin Zick, Joel Heuer, Eike Taegener, Abd Alah Fahed, Agns Voisard, and Joachim W. Fluhr. mHealthAtlas - An expert-based multi-sided platform for the evaluation of mHealth applications. In *2021 IEEE 9th International Conference on Healthcare Informatics (ICHI)*, pages 449-450, 2021.

[Zhou *et al.*, 2020] Yiwu Zhou, Yanqi He, Huan Yang, He Yu, Ting Wang, Zhu Chen, Rong Yao and Zongan Liang. Development and validation a nomogram for predicting the risk of severe COVID-19: A multi-center study in Sichuan, China. *PloS one*, 15(5), e0233328, 2020.

[Zhang *et al.*, 2021] Weishan Zhang, Tao Zhou, Qinghua Lu, Xiao Wang, Chunsheng Zhu, Haoyun Sun, Zhipeng Wang, Sin Kit Lo and Fei-Yue Wang. Dynamic-fusion-based federated learning for COVID-19 detection. *IEEE Internet of Things Journal*, 8(21): 15884-15891, 2021.

[Yan *et al.*, 2021] Zengqiang Yan, Jeffry Wicaksana, Zhiwei Wang, Xin Yang and Kwang-Ting Cheng. Variation-aware federated learning with multi-source decentralized medical image data. *IEEE Journal of Biomedical and Health Informatics*, 25(7): 2615-2628, 2021.

[Wang *et al.*, 2020] Shujun Wang, Lequan Yu, Caizi Li, Chi-Wing Fu and Pheng-Ann Heng. Learning from extrinsic and intrinsic supervisions for domain generalization. In *European Conference on Computer Vision - ECCV 2020*, pages 159–176, 2020.

[Gu *et al.*, 2021] Ran Gu, Jingyang Zhang, Rui Huang, Wenhui Lei, Guotai Wang and Shaoting Zhang. Domain composition and attention for unseen-domain generalizable medical image segmentation. In *Medical Image Computing and Computer Assisted Intervention – MICCAI 2021*, pages 241-250, 2021.

[Hu *et al.*, 2021] Shishuai Hu, Zehui Liao, Jianpeng Zhang and Yong Xia. Domain and content adaptive convolution based multi-source domain generalization for medical image segmentation. *IEEE Transactions on Medical Imaging*, 42(1): 233-244, 2021.

[Pan *et al.*, 2018] Xingang Pan, Ping Luo, Jianping Shi and Xiaoou Tang. Two at once: Enhancing learning and generalization capacities via IBN-Net. In *European Conference on Computer Vision - ECCV 2018*, pages 484–500, 2018.

[Yang *et al.*, 2018] Xin Yang, Haoran Dou, Ran Li, Xu Wang, Cheng Bian, Shengli Li, Dong Ni and Pheng-Ann Heng. Generalizing deep models for ultrasound image segmentation. In *Medical Image Computing and Computer Assisted Intervention – MICCAI 2018*, pages 497-505, 2018.

[Liu *et al.*, 2021] Zhendong Liu, Xiaoqiong Huang, Xin Yang, Rui Gao, Rui Li, Yuanji Zhang, Yankai Huang, Guangquan Zhou, Yi Xiong, Alejandro F Frangi and Dong Ni. Generalize ultrasound image segmentation via instant and plug & play style transfer. In *2021 IEEE 18th International Symposium on Biomedical Imaging (ISBI)*, pages 419-423, 2021.

[Zhang *et al.*, 2020] Ling Zhang, Xiaosong Wang, Dong Yang, Thomas Sanford, Stephanie Harmon, Baris Turkbey, Bradford J. Wood, Holger Roth, Andriy Myronenko, Daguang Xu and Ziyue Xu. Generalizing deep learning for medical image segmentation to unseen domains via deep stacked transformation. *IEEE Transactions on Medical Imaging*, 39(7): 2531 - 2540, 2020.

[Deng *et al.*, 2021] Zhongying Deng, Kaiyang Zhou, Yongxin Yang and Tao Xiang. Domain attention consistency for multi-source domain adaptation. In *British Machine Vision Conference (BMVC)*, pages 4, 2021.

[Wang *et al.*, 2020] Yuxi Wang, Zhaoxiang Zhang, Wangli Hao and Chunfeng Song. Attention guided multiple source and target domain adaptation. *IEEE Transactions on Image Processing*, 30: 892–906, 2020.

[Liu *et al.*, 2021] Quande Liu, Cheng Chen, Jing Qin, Qi Dou and Pheng-Ann Heng. FedDG: Federated domain generalization on medical image segmentation via episodic learning in continuous frequency space. In *2021 IEEE/CVF Conference on Computer Vision and Pattern Recognition (CVPR)*, pages 1013-1023, 2021.

[Zhang *et al.*, 2021] Liling Zhang, Xinyu Lei, Yichun Shi, Hongyu Huang and Chao Cn. Federated learning with domain generalization. *arXiv preprint arXiv:211110487, 2021*.

[Li *et al.*, 2021] Qinbin Li, Bingsheng He and Dawn Song. Model-contrastive federated learning. In *2021 IEEE/CVF*

*Conference on Computer Vision and Pattern Recognition (CVPR)*, pages 10708-10717, 2021.

[Fallah *et al*., 2020] Alireza Fallah, Aryan Mokhtari and Asuman Ozdaglar. Personalized federated learning with theoretical guarantees: A model-agnostic meta-learning approach. In *Neural Information Processing Systems – NeurIPS 2020*, pages 3557-3568, 2020.

[Ulyanov *et al*., 2016] Dmitry Ulyanov, Andrea Vedaldi and Victor Lempitsky. Instance normalization: The missing ingredient for fast stylization. *arXiv preprint arXiv:160708022*, 2016.

[Huang and Belongie, 2017] Xun Huang and Serge Belongie. Arbitrary style transfer in real-time with adaptive instance normalization. In *2017 IEEE International Conference on Computer Vision (ICCV)*, pages 1510-1519, 2017.

[He *et al*., 2016] Kaiming He, Xiangyu Zhang, Shaoqing Ren and Jian Sun. Deep residual learning for image recognition. In *2016 IEEE Conference on Computer Vision and Pattern Recognition (CVPR)*, pages 770-778, 2016.

[Liu *et al*., 2018] Quande Liu, Qi Dou and Pheng-Ann Heng. Shape-aware meta-learning for generalizing prostate MRI segmentation to unseen domains. In *Medical Image Computing and Computer Assisted Intervention – MICCAI 2020*, pages 475-485, 2020.

[Wang *et al*., 2018] Shujun Wang; Lequan Yu; Kang Li; Xin Yang; Chi-Wing Fu; Pheng-Ann Heng. Dofe: Domain-oriented feature embedding for generalizable fundus image segmentation on unseen datasets. *IEEE Transactions on Medical Imaging*, 39(12): 4237-4248, 2020.

[Sivaswamy *et al*., 2015] Jayanthi Sivaswamy, Subbaiah Subbaiah Ramasamy Krishnadas, Arunava Chakravarty, Gopal Datt Joshi and Ujjwal. A comprehensive retinal image dataset for the assessment of glaucoma from the optic nerve head analysis. J*SM Biomed Imaging Data Pap*, 2, 2015.

[Fumero *et al*., 2011] F. Fumero, S. Alayon, J. L. Sanchez, J. Sigut and M. Gonzalez-Hernandez. RIM-ONE: An open retinal image database for optic nerve evaluation. In *2011 24th International Symposium on Computer-Based Medical Systems (CBMS)*, pages 1-6, 2011.

[Orlando *et al*., 2020] José Ignacio Orlando, Huazhu Fu, João Barbosa Breda, Karel van Keer, Deepti R Bathula, Andrés Diaz-Pinto, Ruogu Fang, Pheng-Ann Heng, Jeyoung Kim, JoonHo Lee, Joonseok Lee, Xiaoxiao Li, Peng Liu, Shuai Lu, Balamurali Murugesan, Valery Naranjo, Sai Samarth R Phaye, Sharath M Shankaranarayana, Apoorva Sikka, Jaemin Son, Anton van den Hengel, Shujun Wang, Junyan Wu, Zifeng Wu, Guanghui Xu, Yongli Xu, Pengshuai Yin, Fei Li, Xiulan Zhang, Yanwu Xu and Hrvoje Bogunović. Refuge challenge: A unified framework for evaluating automated methods for glaucoma assessment from fundus photographs. *Medical image analysis*, 59: 101570, 2020.

[McMahan *et al*., 2017] Brendan McMahan, Eider Moore, Daniel Ramage, Seth Hampson and Blaise Aguera y Arcas. Communication-efficient learning of deep networks from decentralized data. In *Proceedings of the 20th International Conference on Artificial Intelligence and Statistics, PMLR*, pages 1273-1282, 2017.